\documentclass[prl,aps,showpacs,twocolumn]{revtex4}

\usepackage{amsmath,amssymb}
\usepackage{graphicx}
\usepackage{psfrag}
\usepackage{graphicx}
\usepackage{epsf}

\newcommand{\signum}{{\rm sgn}} 
\newcommand{\Dcross}{D_\times}
\newcommand{\Dcrossa}{{\tilde D}_\times}
\newcommand{\Dcrossij}{D^{\times}_{ij}}
\newcommand{\Dcrossji}{D^{\times}_{ji}}
\newcommand{\DcrossNullij}{{\Dcrossij}^{(0)}}
\newcommand{\dcross}{d_\times} 
\newcommand{\Ccross}{C_\times}
\newcommand{\Ccrossij}{C^{\times}_{ij}}
\newcommand{\betacross}{N}
\newcommand{\betacrossas}{\tilde{N}}

\begin{document}

\title{Novel universality classes of coupled driven diffusive
  systems}

\author{Abhik Basu$^{1,2}$ and Erwin Frey$^{2,3}$} 

\affiliation{$^{1}$ Poornaprajna Institute of Scientific Research,
  Bangalore, India, \\
  $^2$Abteilung Theorie, Hahn-Meitner-Institut, Glienicker Strasse
  100, D-14109 Berlin, Germany, \\ $^3$Fachbereich Physik, Freie
  Universit\"at Berlin, Arnimalle 14, D-14195 Berlin, Germany }

\date{\today}

\begin{abstract} 
  Motivated by the phenomenologies of dynamic
  roughening of strings in random media and magnetohydrodynamics, 
  we examine the universal
  properties of driven diffusive system with coupled fields. We
  demonstrate that cross-correlations between the fields lead to
  amplitude-ratios and scaling exponents varying continuosly with the
  strength of these cross-correlations. The implications of these
  results for experimentally relevant systems are discussed.
\end{abstract} 


\maketitle
 
Recently significant advances have been made in classifying the
physics of non-equilibrium systems at long time and length scales into
universality classes. It has been shown that standard universality
classes in critical dynamics are quite robust to detailed-balance
violating perturbations~\cite{tauber-etal:02}. Novel features are
found only for models with conserved order parameter and spatially
anisotropic noise correlations. In contrast, truly non-equilibrium
dynamic phenomena, whose steady state can not be described in terms of
a Gibbsian distribution, are found to be rather sensitive to all kinds
of perturbations. Prominent examples are driven diffusive
systems~\cite{schmittmann_zia:review} and diffusion-limited
reactions~\cite{tauber:review}.  For example, one finds that for the
Kardar-Parisi-Zhang equation anisotropic perturbations are relevant in
$d > 2$ spatial dimensions, leading to rich phenomena that include
novel universality classes and the possibility of first-order phase
transitions and multicritical behavior~\cite{tauber_frey:02}.

In this letter we study driven non-equilibrium processes described by
a set of dynamic variables whose dynamics is given in terms of coupled
Langevin equations. Prominent examples include the dynamic roughening
of strings moving in random media \cite{ertacs_kardar:92}, sedimenting
colloidal suspensions \cite{levine-etal:98} and crystals
\cite{lahiri_ramaswamy:97}, and magneto-hydrodynamics
(MHD)~\cite{jkb}. Our goal is to investigate and elucidate some of the
dramatic effects of symmetries of correlation functions on the
universal properties of such systems. We focus on models with two
vector fields, ${\bf u} ({\bf x},t)$ and ${\bf b} ({\bf x},t) $, as
hydrodyamic variables. The quantities of interest are the two
auto-correlation functions, $C^{u}_{ij} ({\bf x},t)= \langle u_i ({\bf
  x},t)u_j({\bf 0},0)\rangle$ and $C^{b}_{ij}({\bf x},t) = \langle
b_i({\bf x},t)b_j({\bf 0},0) \rangle$, and the cross-correlation
function $\Ccrossij ({\bf x},t) = \langle u_i({\bf x},t) b_j({\bf
  0},0)\rangle$; indices $i,j$ refer to cartesian coordinates.  All
these quantities are tensors, whose symmetry properties depend on the
model under consideration.  We are interested in
systems with translational and rotational symmetry, and inversion
symmetry such that $\bf u$ is a polar and $\bf b$ is an axial vector.

In the first part of the letter, we will consider a
one-dimensional Burgers-like model~\cite{jkb} of magneto-hydrodynamics
and its $d$-dimensional generalization~\cite{decay}
\begin{equation}
\frac{\partial {\bf u}}{\partial t} + \frac{\lambda_1}{2}{\nabla}u^2
 + \frac{\lambda_2}{2} {\nabla}b^2= \nu{\nabla^2}{\bf u} +{\bf f},
\label{eq:burgers_1}
\end{equation}
\begin{equation}
\frac{\partial {\bf b}}{\partial t} + 
\lambda_3{\nabla}({\bf u}\cdot{\bf b}) = 
\mu{\nabla^2\bf b} +{\bf g}.
\label{eq:burgers_2}
\end{equation}
Here $\lambda_i$ are coupling constants, $\nu$ and $\mu$ are the
dissipation coefficients, and $\bf f$ and $\bf g$ are external
stochastic forcing functions. These equations are simplified versions
of the dynamical equations governing the time evolution of the
velocity $\bf u$ and the magnetic field $\bf b$ in a magnetized fluid (MHD).
They are constructed in the same spirit as Burgers equation from the
Navier-Stokes equation. In the second part of the letter we will
discuss the advection of a passive vector $\bf b$, where
$\lambda_1=\lambda_2=0$.
   The simplicity of such a model will allow us to explore higher
order correlation functions.

For Langevin equations describing processes relaxing towards a thermal
equilibrium state the correlation functions for the noise have to obey
detailed balance conditions. In non-equilibrium models there are no
such restrictions. As a minimal requirement one might ask that the
noise terms $\bf f$ and $\bf g$ in the Langevin equations obey the
same symmetries as the correlation functions for the hydrodynamic
fields. Since ${\bf u}$ is a polar vector and $\bf b$ is an axial
vector, $\langle u_i({\bf k},t)u_j({\bf -k},0)\rangle,\, \langle
b_i({\bf k},t)b_j({\bf -k},0) \rangle$ are real and even in ${\bf k}$,
but the cross-correlation function $\Ccrossij ({\bf k},t)=\langle
u_i({\bf k},t) b_j({-\bf k},0)\rangle$ is imaginary and odd in $\bf
k$~\cite{jkb}. Then, assuming Gaussian distributed conserved noise
with zero mean, the noise correlation functions have to be of the
following form
\begin{eqnarray}
  \langle f_i ({\bf k},t) f_j ({-{\bf k},0}) \rangle 
          &=&  2 k_i k_j D_u^{(0)} ({\bf k}) \delta (t) \\  
  \langle g_i ({\bf k},t) g_j ({-{\bf k},0}) \rangle 
          &=&  2 k_i k_j D_b^{(0)} ({\bf k}) \delta (t) \\  
  \langle f_i ({\bf k},t) g_j ({-{\bf k},0}) \rangle 
          &=&  2 i {\DcrossNullij} ({\bf k}) \delta (t) 
\end{eqnarray}
where the noise variances $D_{u,b}^{(0)} ({\bf k})$ are even and
${\DcrossNullij} ({\bf k})$ is odd in $\bf k$, respectively.
Equations (3) and (4) are invariant under inversion, rotation and
exchange of $i$ with $j$. We take the noise cross-correlation, Eq.\ (5)
to be invariant under inversion, but we allow it to break
rotational invariance or symmetry with respect to an interchange of
the cartesian indices $i$ and $j$.

We are interested in the physics at long time and length scales. Then
all the correlation functions $C({\bf x},t)$ are expected to obey
scaling relations of the form
\begin{eqnarray}
 C ({\bf x},t) = x^{2 \chi} C (t/x^z).
 \label{eq:scaling}
\end{eqnarray}
Since we have two independent fields $\bf u$ and $\bf b$ there could
in principle be two different roughness exponents $\chi_{u,b}$. Due to
Galilean invariance, however, none of the non-linearities in the
equations of motion renormalize, and one gets $\chi_u = \chi_b = \chi
= 2 - z$ \cite{jkb,fns}.

\paragraph{Symmetric Cross-Correlations. --} If both the fields are 
irrotational, one can introduce two scalar fields $h$ and $\phi$ such
that ${\bf u} = \nabla h$ and ${\bf b} = \nabla \phi$; note that
$\phi$ is actually a pseudo-scalar. Then
Eqs.(\ref{eq:burgers_1},\ref{eq:burgers_2}) become identical to a
model of Erta\c{s} and Kardar \cite{ertacs_kardar:92} describing the
dynamic roughening of directed lines,
\begin{equation}
{\partial h\over\partial t}+{\lambda_1\over 2}(\nabla h)^2
+{\lambda_2\over 2}(\nabla \phi)^2=\nu\nabla^2h+\eta_h \, ,
\label{eq:ertacs_kardar_1}
\end{equation}
\begin{equation}
{\partial \phi\over\partial t}
+\lambda_3(\nabla h) (\nabla\phi)=\mu\nabla^2\phi
+\eta_\phi \, ,
\label{eq:ertacs_kardar_2}
\end{equation}
where ${\bf f} = \nabla \eta_h$ and ${\bf g} = \nabla \eta_{\phi}$.
The cross-correlation function $\DcrossNullij$ is now symmetric in the
tensor indices 
and $\langle h({\bf k},0)\phi(-{\bf
  k},0)\rangle$ is imaginary and odd in $\bf k$. If, in addition, we
require rotational invariance, the cross-correlation function would
vanish. This is the case considered in Ref.\cite{ertacs_kardar:92}. For a
truley non-equilibrium model there is, however, no physical principle
which would exclude a finite cross-correlation term a priori. Hence we
allow for a non-zero $\langle \eta_h ({\bf k},0) \eta_\phi (-{\bf
  k},0)\rangle$, which then explicitely breaks rotational invariance,
and explore its consequences for the dynamics.

We have determined the roughness exponent $\chi$ and the dynamic
exponent $z$ employing a lowest order self-consistent mode coupling
scheme and a one-loop dynamic renormalisation group calculation.
Perturbation theory is formulated in terms of the response and
correlation functions for the fields $h$ and $\phi$. They are
conveniently written in terms of self-energies $\Sigma (k,\omega)$ and
generalized kinetic coefficients $D (k,\omega)$.  For
simplicity we assume that $\nu=\mu$; in MHD this
would correspond to a system with magnetic Prandtl number
$P_m=\mu/\nu=1$.  Then there is only one response function and it can
be written as $G_{h,\phi}^{-1}({\bf k},\omega) = i \omega - \Sigma
({\bf k},\omega)$. Then, correlation functions are of the form,
$C_\alpha ({\bf k},\omega) = 2 D_\alpha ({\bf k},\omega) | G ({\bf
  k},\omega)|^2$ for $\alpha = h, \phi$ and $\Ccross ({\bf k}, \omega)
= 2i \Dcross ({\bf k},\omega) | G ({\bf k},\omega)|^2$ for the
cross-correlation function. In diagrammatic language lowest order 
mode-coupling theory is equivalent to a self-consistent one loop theory.
The ensuing coupled set of integral equations is compatible with the
scaling form Eq.~\ref{eq:scaling}. In Fourier space the scaling form
reads for the self energy, $\Sigma (k, \omega) = \Gamma k^z \sigma
(\omega/k^z)$, and for the generalized kinetic coefficients $D_h
(k,\omega) = D_h k^{-d-2\chi} d_h(\omega/k^z)$, $D_{\phi} (k,\omega) =
D_\phi k^{-d-2\chi} d_\phi (\omega/k^z)$, $\Dcross ({\bf k},\omega)=
\signum ({\bf k}) \Dcross k^{-d-2\chi} \dcross (\omega/k^z)$. To solve
this set of coupled integral equations we employ a small
$\chi$-expansion~\cite{jkb2}.  This requires matching of the self
energies and correlation functions at $\omega=0$.  With the
zero-frequency expressions $\Sigma({\bf k},0) = \Gamma\,k^z$, $D_h
({\bf k},0) = D_h k^{-2\chi-d}$, $D_{\phi} ({\bf k},0) = D_\phi
k^{-2\chi-d}$, one finds for the one-loop self-energy (we take $\lambda_1= 
\lambda_2=\lambda_3=\lambda$ without any loss of generality)
\begin{equation}
 \frac{\Gamma^2}{D_h\lambda^2} = \frac{S_d}{(2\pi)^d} 
 \frac{1}{2d} \left(1+\frac{D_\phi}{D_h}\right) \, ,
\label{selfen}
\end{equation}
and for the one-loop correlation functions, 
\begin{eqnarray}
 \frac{\Gamma^2}{D_h \lambda^2} 
 &=& \frac14 \frac{S_d}{(2\pi)^d} 
     \frac{1}{d-2+3\chi}
     \left[ 1 + \left(\frac{D_\phi}{D_h}\right)^2 + 
               2\left( \frac{\Dcross}{D_h}\right)^2
     \right], \nonumber \\
 \frac{\Gamma^2}{D_\phi \lambda^2}
 &=& \frac12 \frac{S_d}{(2\pi)^d} \frac{1}{d-2+3\chi}
     \left[ \frac{D_h}{D_\phi} - 
            \left( \frac{{\Dcross}}{D_\phi}\right)^2
     \right].
\label{selfcr}
\end{eqnarray}
Here $S_d$ is the surface of a $d$-dimensional unit sphere.  From
Eqs.(\ref{selfcr}) we find
\begin{equation}
  \left(\frac{D_\phi}{D_h}\right)^2 + 
  2 \betacross \left( \frac{D_h}{D_\phi} + 1 \right) - 1 = 0,
\label{d1byd2}
\end{equation}
where $\betacross \equiv ({\Dcross / D_h})^2$ defines an amplitude
ratio. In the Eq.(\ref{d1byd2}), the domain of $N$ is determined by the range of
real values for $D_{\phi}/D_h$ starting from 1 (for $N=0$). 
Thus for small $\betacross$ we can expand around 0 and
look for solutions 
 of the form ${D_\phi / D_h} = 1 + a\betacross$, such that for
$\betacross=0$ we recover $D_h=D_\phi$ (the result of
Ref.~\cite{ertacs_kardar:92}).  We obtain $a=-2$, i.e.,
\begin{equation}
D_\phi/D_h = 1-2\betacross \, ,
\label{ampratio}
\end{equation}
implying that within this approximate calculation $\betacross$ cannot
exceed $1/2$, i.e., $\Dcross \leq D_h/2$. An important consequence of
this calculation is that the amplitude ratio $D_\phi / D_h$ is no longer
fixed to $1$ but can vary continuously with the strength of the noise
cross-correlation amplitude $\Dcross$.  These results are confirmed by
a one-loop RG for the strong coupling fixed point in $d=1$. In
addition, Eq. (12) is valid at the roughening transitions to lowest
order in a $d=2+\epsilon$ expansion.

In contrast, the scaling exponents $\chi$ and $z$ are not affected by
the presence cross-correlations. We get $\chi=\frac12$ and $
z=\frac32$ in $d=1$ dimensions, $\chi=-O(\epsilon)^2$ and
$z=2+O(\epsilon)^2$ at the roughening transitions in a $d=2+\epsilon$
expansion, and $\chi=2-{d\over 2}$ and $z={4\over 3}+{d\over 6}$ for
the strong coupling phase. Note that the values for the strong
coupling exponents are obtained within Bhattacharjee's~\cite{jkb2}
small-$\chi$ expansion, as described above. There is still an ongoing
debate whether those values for the exponents actually correspond to
the KPZ strong coupling case (see Ref.\cite{frey1} for a discussion).

We have verified our analytical results in $d=1$ by numerical
simulations of both a coupled lattice model with cross-correlations, 
and direct numerical simulations of the model equations
(\ref{eq:burgers_1}) and (\ref{eq:burgers_2}).  Our numerical results
explicitly demonstrate the dependence of the amplitude-ratio on the
cross-correlation function amplitude.  Details will be published
elsewhere \cite{abfrey}.

\paragraph{Anti-Symmetric Cross-Correlations.--} 
In the preceding paragraph we
have restricted ourselves to irrotational fields.
If the vector fields $\bf a= u,\,b$ have the form ${\bf a}=\nabla
\times {\bf V}_a + \nabla S_a,$ with vectors ${\bf V}_a$ being
cross-correlated but scalars $S_a$ uncorrelated then the variance $\Dcrossij$
satisfies $\Dcrossij ({\bf
k})
= -\Dcrossij ({\bf -k}) = \Dcrossji ({\bf -k}) = -[ \Dcrossij ({\bf k})]^*.$.
This is the antisymmetric part of the crosscorrelations.
The noise strength $\Dcrossa$ is defined by $\Dcrossij
({\bf k}) \Dcrossji ({\bf -k})= \Dcrossa^2 k^4$. In the scaling limit,
the self energy reads $\Sigma(k,\omega)= \Gamma
k^z\sigma(\omega/k^z)$, the correlation functions are $C^u_{ij}
(k,\omega) = k_i k_j D_u k^{-d-2\chi} \, d_u(\omega/k^z)$, $C^b_{ij}
(k,\omega) = k_i k_j D_b k^{-d-2\chi}\, d_b(\omega/k^z)$, and the
antisymmetric part of the crosscorrelation function reads $C_{ij}^a
(k,\omega) = D_{ij}^a ({\bf k})k^{-2\chi-z-d}$.

Following methods used for the symmetric crosscorrelations, we obtain
the analogues of Eqs.(\ref{selfen}) and (\ref{selfcr})
\begin{equation}
 \frac{\Gamma^2}{D_u\lambda^2} = \frac{S_d}{(2\pi)^d}
 \frac{1}{2d} \left(1+\frac{D_b}{D_u}\right) \, ,
\label{selfena}
\end{equation}
\begin{eqnarray}
 \frac{\Gamma^2}{D_u \lambda^2}
 &=& \frac14 \frac{S_d}{(2\pi)^d}
     \frac{1}{d-2+3\chi}
     \left[ 1 + \left(\frac{D_b}{D_u}\right)^2 +
               2\left( \frac{\Dcrossa}{D_u}\right)^2
     \right], \nonumber \\
 \frac{\Gamma^2}{D_b \lambda^2}
 &=& \frac12 \frac{S_d}{(2\pi)^d} \frac{1}{d-2+3\chi}
     \left[ \frac{D_u}{D_b} +
            \left( \frac{{\Dcrossa}}{D_b}\right)^2
     \right].
\label{selfcra}
\end{eqnarray}
Equations (\ref{selfena}) and (\ref{selfcra}) give $D_u/D_b=1$ at the fixed
point for arbitrary values of $\betacrossas=({\Dcrossa / D_h})^2$.
Hence no restrictions on $\betacrossas$ arises from that.  In contrast to the
effects of the symmetric crosscorrelations,
the exponents now depend continuously on $\betacrossas$.
To leading order, we get
\begin{equation}
\chi= \frac{2}{3}- \frac{d}{6}+ \frac{\betacrossas d}{6} \, , 
\quad
z=\frac{4}{3}+\frac{d}{6}-\frac{\betacrossas d}{6}.
\label{expo}
\end{equation}
These exponents presumably describe the rough phase above $d>2$, with
the same caveats as above~\cite{frey1}.  With increasing $\Dcrossa$
the exponent $\chi$ also grows (and $z$ decreases). Obviously this
cannot happen indefinitely. We estimate the upper limit of
$\betacrossas$ in the following way: Notice that the
Eqs.(\ref{eq:burgers_1}) and (\ref{eq:burgers_2}) along with the
prescribed noise correlations (i.e., equivalently the dynamic
generating functional) are of conservation law form, i.e.\/ they
vanish as $\bf k\rightarrow 0$.  Thus there is no information of any
infrared cut off in the dynamic generating functional.  Moreover, we
know the solutions of the equations {\em exactly} if we drop the
non-linear terms (and hence, the exponents: $\chi=1-d/2,\,z=2$). Hence
physically relevant quantities like the total kinetic and magnetic
energies, $\int_k\langle {\bf u} ({\bf k},t) {\bf u} (-{\bf
  k},t)\rangle$ and $\int_k\langle {\bf b} ({\bf k},t) {\bf b} (-{\bf
  k},t) \rangle$, remain {\em finite} as the system size diverges, and
are thus independent of the system size.  Since the non-linear terms
are of the conservation law form, inclusion of them {\em cannot }
bring a system size dependence on the values of the total kinetic and
magnetic energies. However, if $\chi$ continues to increase with
$\Dcrossa$ at some stage these energies would start to depend on the system
size which is unphysical \cite{foot1}. So we have to restrict
$\betacrossas$ to values
smaller than the maximum value for which these energy integrals
are just system size independent: This gives
$\betacrossas^\text{max}={2\over d}(d/2+1)$.  Note that the limits on
$\betacross$ and $\betacrossas$ impose consistency conditions on the
amplitudes of the measured correlation functions but not on the bare
noise correlators.  

Antisymmetric cross-correlations stabilise the
short-range fixed point with respect to perturbations by long-range
noise with correlations $\propto k^{-y},\, y>0$.  This can easily be
seen: In presence of noise correlations sufficiently singular in the
infra-red limit, i.e.\/ large enough $y$, the dynamic exponent is {\em
  exactly} given by \cite{amit,frey1} $z_\text{lr} = \frac{2+d}{3} -
\frac{y}{3}$.  The short range fixed point remains stable as long as
$z_\text{sr}<z_\text{lr}$ which gives $y<-2+(1+\betacrossas)d/2$.
Hence we conclude that in presence of antisymmetric cross-correlations
long range noise must be {\em more singular} for the short range noise
fixed point to loose its stability or in other words, antisymmetric
crosscorrelations increases the stability of the of the short range noise fixed
point with respect to perturbations from long range noises.

We have seen that the amplitudes of the 
cross-correlation function play a quite crucial role in determining
the long wavelength properties of the system. In our analysis we used
only short range noise, which is enough to elucidate the basic points.
However, a Langevin description of many systems often requires a
noise term with correlations becoming singular in the long wavelength
limit, such as fully developed MHD \cite{jkb}. These systems are
typically characterized by a set of anomalous exponents for higher
order correlation functions. Below we give an illustrative example to
highlight the effects of symmetries on the anomalous scaling exponents
of higher order correlation functions in the passive vector limit 
where the velocity field $\bf u$ is
assumed to obey a Gaussian distribution (instead of Eq.(\ref{eq:burgers_1}))
with a variance $\langle
u_i({\bf k},t) u_j({\bf -k},0)\rangle= \frac{2D \delta (t)}
{(k^2+M^2)^{d/2+\epsilon/2}}[\alpha P_{ij}+ Q_{ij}]$ where $0\leq
\epsilon \leq 2$, which makes the model analytically tractable. As
before, the magnetic field $\bf b$ is governed by
Eq.(\ref{eq:burgers_2}).  The tensor $P_{ij}$ is the transverse
projection operator, $Q_{ij}$ is the longitudinal projection operator.
The parameter $\alpha >0$ determines the extent of incompressibility
of the $\bf u$ field. Thus in this problem $\alpha$ appears as a
tuning parameter in the multiplicative noise,
very much like $\betacross$ and $\betacrossas$
appeared in our previous results. By following a field-theoretic
dynamic renormalization group procedure in conjunction with Operator
Product Expansion \cite{adm}, we calculate the scaling exponents of
the structure functions $S_n(r)=\langle [\theta ({\bf x+r})-\theta
({\bf x})]^{2n}\rangle\sim r^{\zeta_n},\; {\bf b}=\nabla \theta$.
Within a one-loop approximation we find
\begin{eqnarray}
\zeta_n = 
2n &-& \frac{n\epsilon d} {d\alpha +1-\alpha}
    \biggl[ \alpha + \frac{{1-\alpha}}{d} \nonumber \\
                  &+&  2(n-1) \left\{ \frac{\alpha} {d} + 
                                      \frac{3(1-\alpha)} {d(d+2)}
                                \right\}
    \biggr] \, .
\label{anoexp}
\end{eqnarray}
This clearly demonstrates that even for the linear problem there is an
continuous dependence of the scaling exponents on the parameter
$\alpha$, characterizing the extent to which the velocity field is
compressible. We expect this to hold also for the nonlinear problem,
whose analysis is significantly more complicated.

Let us now review our results in the
context of some some physically relevant systems.  
Our results are relevant for a wide class of non-equilibrium systems.
In MHD turbulence
the cross-correlation function $\langle u_i({\bf k},t)b_j({\bf
  -k},t)\rangle$ is, in general, nonzero~\cite{thesis}, 
and as before, is odd and imaginary in $\bf k$.  
Similar calculations as here for MHD show that two
dimensionless numbers, the magnetic Prandtl number 
$P_m$ and the ratio of the magnetic to the kinetic
energy, are non-universal; they are functions of both the symmetric
and antisymmetric part of the cross-correlation amplitudes.  Another
system of interest is the dynamics of a drifting polymer
through a solution~\cite{er}. Here the hydrodynamic
degrees of freedom are the transverse and longitudinal displacements
with respect to the mean position.  Dynamic light scattering
experiments can be preformed to investigate the effects of
cross-correlations discussed here. Our results are significant also for 
coupled growth of nonequilibrium surfaces
\cite{basi}, and sedimenting lattices \cite{lahiri_ramaswamy:97,dib}.

In summary, we have demonstrated that cross-correlations between two
vector fields can drastically alter their asymptotic statics and
dynamics at long length and time scales. The symmetric and
anti-symmetric part of the noise cross-correlation function have
different effects. The symmetric part leaves the scaling exponents
unaffected but yields amplitude ratios of the various correlation
functions, which continuously depend on the amplitude of the noise
cross-correlation (see Eq.(12)). In contrast, the asymmetric part
leaves the amplitude ratios unaffected, but leads to continuously
varying exponents (see Eq.(15)). In both cases the continuos variation
with the noise amplitude of the cross-correlations is not arbitray but
constrained by scaling relations (see Eqs.(12) and (15)), a feature, present 
also in our results on the multiplicative noise driven linear system. We have
shown this using renormalization group methods and 
mode coupling theory, confirmed by some preliminary
simulations~\cite{abfrey}. Recently, Drossel and Kardar~\cite{barbara}
have studied a set of coupled Langevin equations describing the
interplay between phase ordering dynamics in the bulk and roughening
dynamics of the interface of binary films. They find a similar
continuous variation of the dynamical exponent with the coupling
strength of the bulk and surface fields.
Nonperturbative analysis or numerical simulations may be necessary to
resolve the questions about the rough phase more satisfactorily. In
the light of our results it might also be interesting to examine the
effects of cross-correlations on the multiscaling properties of MHD in
experiments and/or numerical simulations.

One of the authors (AB) wishes to thank the Alexander von Humboldt
Stiftung, Germany for financial support.  We thank
S.~Ramaswamy, J.K.~Bhattacharjee and S.M.~Bhattacharjee for useful
discussions.


\begin{thebibliography}{99}
  
\bibitem{tauber-etal:02} U.C. T\"auber {\em et al.\/}, Phys. Rev.
  Lett. {\bf 88}, 045701 (2002).
  
\bibitem{schmittmann_zia:review} B. Schmittman and R.K.P. Zia, in {\it
    Phase transitions and critical phenomena}, eds. C. Domb and J.L.
  Lebowitz, Vol. 17 (Academic Press, London, 1995).
  
\bibitem{tauber:review} U.C. T\"auber, Adv. in Solid State Physics
  {\bf 43}, ?? (2003)
  
\bibitem{tauber_frey:02} U.C. T\"auber and E. Frey, Europhys. Lett.
  {\bf 59}, 655 (2002).
  
\bibitem{ertacs_kardar:92} D. Erta\c{s} and M. Kardar, Phys.  Rev.
  Lett. {\bf 69}, 929 (1992).
 
\bibitem{levine-etal:98} A. Levine et al., Phys. Rev. Lett. {\bf 81},
  5944 (1998).
 
\bibitem{lahiri_ramaswamy:97} R. Lahiri and S. Ramaswamy, Phys.  Rev.
  Lett. {\bf 79}, 1150 (1997).
  
\bibitem{jkb} S. Yanase, Phys. Plasma {\bf 4}, 1010 (1997); J.
  Fleicher and P.H. Diamond, Phys. Rev. E {\bf 58}, R2709 (1998); A.
  Basu, J.K. Bhattacharjee and S. Ramaswamy, Eur. Phys. J. B {\bf
    9}, 725 (1999).

\bibitem{fns}D. Forster, D. R. Nelson, and M. J. Stephen, {\em Phys. Rev A},
{\bf 16}, 732 (1977).
  

\bibitem{decay} A. Basu, Phys. Rev. E {\bf 61}, 1407 (2000).

\bibitem{jkb2} J.K. Bhattacharjee, J. Phys. A, {\bf 31},
  {L93} (1998).
  
\bibitem{frey1} E. Frey, U.C. T\"{a}uber, and H.K. Janssen, Europhys.
  Lett. {\bf 47}, 14 (1999); H.K. Janssen, U.C.  T\"{a}uber and E.
  Frey, Eur. Phys. J B {\bf 9}, 491 (1999).
  
\bibitem{abfrey} A. Basu and E. Frey, unpublished.
  
\bibitem{foot1} In contrast with our model system, physical systems,
  such as fluid and MHD turbulence, where total energy diverges with
  the system size are described by stochastically driven Langevin
  equations with noise correlators diverging with the system size.
  
\bibitem{amit} A.K. Chattopadhyay and J.K. Bhattacharjee, Europhys.
  Lett., {\bf 42}, 7381 (1998).
  
\bibitem{adm} L.T. Adzhemyan {\em et al.\/}, Phys.  Rev. E, {\bf 58}, 1823
  (1998).

\bibitem{thesis} A. Basu, A. Sain, S.K. Dhar and R. Pandit, Phys.
  Rev. Lett. {\bf 19}, 19 (1999); A. Basu, PhD Thesis, Indian
    Institute of Science (1999).
  
\bibitem{er} D. Erta\c{s} and M. Kardar, Phys. Rev. E {\bf 48},
  1228 (1993).
  
\bibitem{barbara} B. Drossel and M. Kardar, cond-mat/0307579 (eprint).
  
\bibitem{basi} A.L. Barabasi, Phys. Rev. A {\bf 46}, R2977 (1992).
  
\bibitem{dib} D. Das {\em et al.\/}, Phys. Rev.  E {\bf 64}, 021402
  (2001).


\end{thebibliography}
\end{document}